# Some Peculiarities of Dielectric Spectroscopy in Ferroelectric Nematics

Yuri P. Panarin,[1,2,*] Neelam Yadav,[1] Rahul Uttam,[1] Wanhe Jiang,[3] Georg H. Mehl,[3] Jagdish K. Vij[1]

*[1]Department of Electronic and Electrical Engineering, Trinity College, Dublin 2, Ireland*
*[2]School of Electrical and Electronic Engineering, TU Dublin, Dublin 7, Ireland*
*[3]Department of Chemistry, University of Hull, Hull HU6 7RX, UK*

Dielectric spectroscopy is known as one of the most powerful techniques for studying ferroelectric and other polar materials. Since the discovery of ferroelectricity in Liquid Crystals, it has been successfully employed for the characterization of ferro-, antiferro- and ferri-electric liquid crystalline phases. However, recently the Boulder group raised the question of the applicability of dielectric spectroscopy for characterizing ferroelectric nematics due to parasitic effects from the insulating alignment layers. This affects the apparent/measured values of the dielectric permittivity. In this paper, we study this effect in greater detail. The following issues will receive special attention: Are the real values of dielectric permittivity lower or higher than the measured ones? Can the real values of dielectric permittivity be recovered in the cell with alignment layers? We also provide an example of the effect of insulating alignment layers in a non-ferroelectric nematic phase.

**\*Corresponding author:** yuri.panarin@tudublin.ie

## I.     INTRODUCTION

Dielectric spectroscopy is one of the most sensitive methods for studying ferroelectric and other polar materials/phases. It has been successfully employed for characterization of ferro- [1,2,3,4] / antiferro- [5,6] and ferri-electric [7,8] liquid crystalline (LC) phases. The technique was also applied for the characterization of two independently discovered compounds exhibiting the ferroelectric nematic (NF) phase [9,10]. Ferroelectric nematic materials are characterized by a large spontaneous electric polarization ~ 5 μC/cm$^2$ due to a strong order of the dipoles, resulting in giant / Colossal dielectric Permittivity (CP) of the

order of 10,000 and even higher [9,11,12,13, 14,15,16,17,18,19,20,21]. CP was also observed earlier in the bent-core LC compounds [22,23].

Recently, N. Clark et al poured a cold shower on the entire dielectric society, raising the question of the applicability of dielectric spectroscopy for ferroelectric nematics due to the effect of insulating alignment layers on the apparent/measured values of the dielectric permittivity [24].They noticed that the apparent capacitance ($C_{ap}$) of LC cell can be expressed as a combination of two capacitances in series: the capacitance of the LC layer ($C_{LC}$) and capacitance of the insulating alignment layers, $C_{al}\left(\frac{A.\varepsilon_0.\varepsilon_{al}}{2d_{al}}\right)$. Hence, the apparent capacitance of LC cell is given as:

$$C_{ap}(\omega) = \frac{C_{LC}(\omega) \cdot C_{al}}{C_{LC}(\omega) + C_{al}} \qquad (1)$$

Here, $C$ is capacitance, $\varepsilon$ is dielectric permittivity, and $d$ is thickness of the LC cell, with subscripts *LC*, *al*, and *ap* indicating liquid crystal, alignment layer, and apparent, respectively.

There can be two possible opposite cases with reference to Eq. (1): the ordinary case, where $C_{LC} \ll C_{al}$, and the extraordinary case, where $C_{LC} \gg C_{al}$. In the ordinary case, i.e., in liquid crystals with low/moderate dielectric permittivity, the capacitance of the LC layer is much smaller than the capacitance of the alignment layer, $C_{LC} \ll C_{al}$, and hence the apparent capacitance according to Eq. (1) equals to capacitance of the LC cell ($C_{ap} = C_{LC}$). This gives the real/actual values of capacitance and dielectric permittivity of the measured sample.

The second, extraordinary scenario occurs in the liquids/liquid crystals with very high dielectric permittivity ($\varepsilon > 10,000$), such as in the ferroelectric nematics and the bent-core compounds [22,23]. In this case, the capacitance of the LC layer exceeds the capacitance of the alignment layer, $C_{LC} \gg C_{al}$. In such extraordinary case, the apparent capacitance is limited to the capacitance of the alignment layers, $C_{al}$ i.e., constrained by Clark's limit, $C_{ap} \lesssim C_{al}$. In terms of dielectric permittivity, the Clark's limit/criterion can be given as:

$$\varepsilon_{ap} = \frac{(d_{LC}+2d_{al}) \cdot \varepsilon_{al}}{d_{al}} \cong \frac{d_{LC} \cdot \varepsilon_{al}}{d_{al}} \qquad (2)$$

The apparent value limit due to alignment layer was studied experimentally, confirming its proportionality to the cell thickness. In other words, in the extraordinary case, the apparent capacitance linearly depends on the ratio $d_{LC}/d_{al}$. This feature was observed and confirmed experimentally [25,26,27,28,29]. However, the actual value of the LC capacitance, even when limited/distorted, can be reconstructed from the apparent value. In the next section, we will discuss different ways to process this reconstruction.

## II. RECONSTRUCTION OF LC CAPACITANCE FROM APPARENT VALUE

The apparent capacitance of sandwich-type liquid crystal cells measured by dielectric spectroscopy depends on both the capacitance of liquid crystal ($C_{LC}$) and the capacitance of the alignment layer ($C_{al}$), as described by Eq. (1) and illustrated in Fig. 1(a). It should be noted that this dependence is rather smooth, so that even at $C_{LC} = 10 \cdot C_{al}$, the apparent capacitance is not completely saturated, reaches only 90% of the limit value $C_{al}$ and still has 10 % to increase, showing a weaker temperature dependence on approaching $C_{ap} \lessapprox C_{al}$.

Hence, the real value of the liquid crystal capacitance $C_{LC}$ (and dielectric permittivity), even when affected by the capacitance of the alignment layer, can be deduced or recovered from the Eq. (1) and is rewritten as:

$$C_{LC}(\omega) = \frac{C_{al} \cdot C_{ap}(\omega)}{C_{al} - C_{ap}(\omega)} \qquad (3)$$

This gives an idea for recovering the real values of the capacitance / dielectric permittivity from the apparent values. The dependence of Eq. (3) is illustrated in Fig. 1(b), which shows the actual capacitance as a function of the apparent capacitance of the LC cell for $C_{al} = 1$.

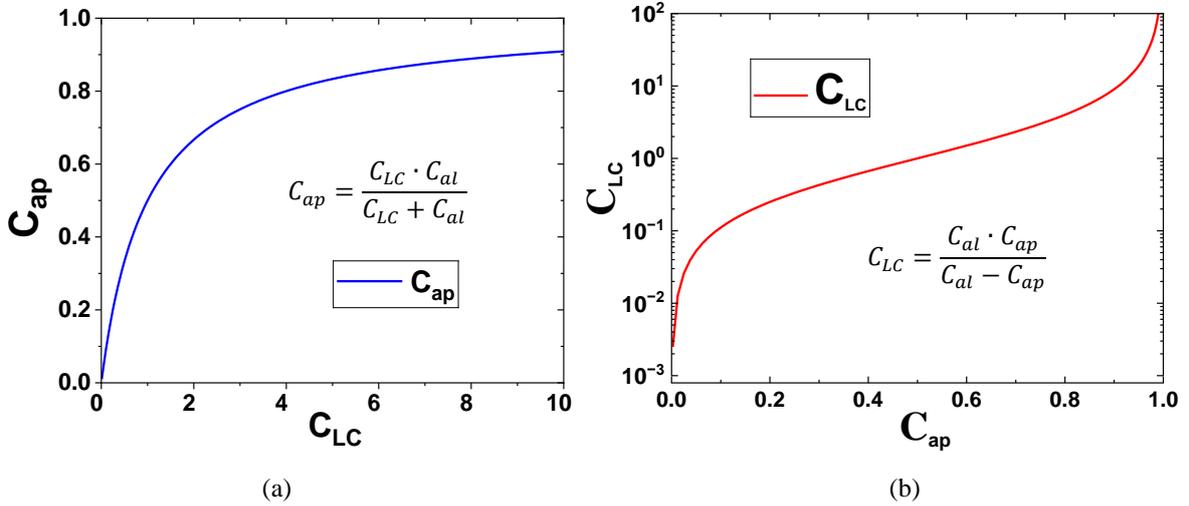

(a)          (b)

FIG 1. Schematical diagram of (a) the dependence of apparent capacitance on the capacitance of the LC layer, and (b) the dependence of recovered capacitance $C_{LC}$ on the apparent capacitance $C_{ap}$ for capacitance of alignment layers $C_{al} = 1$.

Hence, knowing the capacitance of the alignment layers, one can easily convert the measured/apparent capacitance to the capacitance of liquid crystal and restore the true value of the dielectric permittivity. However, as mentioned before, the restoration of the dielectric permittivity depends on several other conditions, namely the ratio between $C_{LC}$ and $C_{al}$. Depending on this ratio, there are two opposite cases: the ordinary one (Case 1), where $C_{LC} \gg C_{al}$, and extraordinary (Case 2), where $C_{LC} \gg C_{al}$. However, our primary interest lies in the intermediate case, which allows us to obtain the real value of the permittivity from the apparent value. Let us briefly discuss these three different cases.

1. Case 1. $C_{LC} \ll C_{al}$ This is an ordinary case where the apparent capacitance is practically the same as the real capacitance of LC layer and hence does not require any correction, or practically similar within the experimental error of 3 %
2. Case 2. $C_{LC} \gg C_{al}$ This is opposite to the Case 1 where the apparent capacitance is almost the same as capacitance of alignment layers, and the Eq. (3) cannot be used straight-forward. However, in some particular case (see below) the actual capacitance can be recovered.
3. Case 3. This is intermediate case which boundaries depend on experimental error. Typical experimental error for $C_{ap}$ ~3 % and assuming this value we define the range of intermediate case as $0.03 \cdot C_{al} < C_{LC} < 30\ C_{al}$. This is the most practical case where the formula Eq. (3) can be used straightforward.

Consequently, the conditions for ordinary and extraordinary cases can be rewritten as: $C_{LC} < 0.03 \cdot C_{al}$ and $C_{LC} > 30\ C_{al}$ correspondingly.

In the following sections we will show some experimental examples of Clarks limit in dielectric measurement in the LC materials with CP and recovery of dielectric permittivity for Cases 2 and 3.

### III. RESULTS

#### A. Materials

Dielectric spectroscopy measurements were performed on the newly synthesized ferroelectric nematic compound WJ-16, with its molecular structure and phase transitions detailed in Fig. 2.

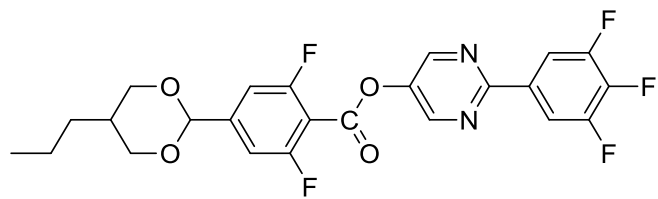

On cooling: Cr 79.3 SmA 110.5 N 198.6 Is

FIG 2. The molecular structure, phase-sequence, and transition temperatures of WJ-16.

This compound was derived from DIO by replacing the fluorinated benzene ring with pyrimidine. This substitution increases the dipole moment by 1 D to 10.8 D. However, this increase suppresses the $N_F$ phase and induces SmA phase instead [30]. The compound does not exhibit any ferroelectric properties; instead, it shows perfect homogeneous, domain-free texture, hysteresis-free switching and linear P-E dependence, i.e. behaves overall as a typical paraelectric [30]. The appearance of ferroelectricity in nematics has been studied theoretically and experimentally. Madhusudana theoretically showed the importance of a specific charge distribution along the molecule for the formation of the ferroelectric nematic phase [31]. Finally, Li *et al* [13] systematically analyzed more than 100 new ferrogenic nematic compounds and defined the Pearson's coefficient (or the impact) of different parameters responsible for the ferroelectric nematic phase. The most important pre-requisites are the dipole moment (0.26); molecular length (0.19); and the dipole angle (0.16). Hence, the dipole moment itself is the most important parameter, but it cannot guarantee the formation of the ferroelectric nematic phase. Therefore, the absence of the $N_F$ phase in WJ-16, which has a higher dipole moment than DIO is not a surprise. This evidences that a large dipole moment is not a sufficient condition for the formation of the ferroelectric nematic phase [13,21].

While being non-ferroelectric, WJ-16, however, exhibits very high or Colossal Permittivity (CP) in the paraelectric phase, allowing this material to be classified as a Superparaelectric (SPE). Further details on the assignment and other features of SPE are given in Ref. [30]. Here, we just use this as an example to show that Clark's limit is observed in non-ferroelectric liquid crystals as well.

### B. Dielectric spectroscopy of the planar LC cells, Cases 1 and 2

#### *1. Dielectric spectroscopy of polymer coated planar cells, Case 2*

Dielectric spectroscopy measurements over a frequency range 0.1 Hz – 10 MHz were made using a broadband Alpha High Resolution Dielectric Analyzer (Novocontrol GmbH, Germany). We used planar commercial cells (E.H.C ltd., Japan) of different thicknesses (2, 4, 9, 15 and 25 μm) with 20 nm polyimide alignment layers and 0.5 cm$^2$ electrodes area. We also used hand-made cells with the uncoated (bare) ITO electrodes to avoid possible from the alignment layers. The ITO electrodes used in these cells have very low sheet resistance (5 Ω/□) to avoid the parasitic peak arising from the sheet resistance of ITO in series with the capacitance of the cell. The measurements were carried out under the application of weak AC voltage of 0.1 V applied across the cell. The temperature of the sample is stabilized to within ± 0.05 °C using Eurotherm 2604 temperature controller. The dielectric spectra are analyzed using the Novocontrol WINFIT program. Figure 3 presents the temperature dependence of dielectric permittivity at 0.1 Hz for planar cells of different thicknesses.

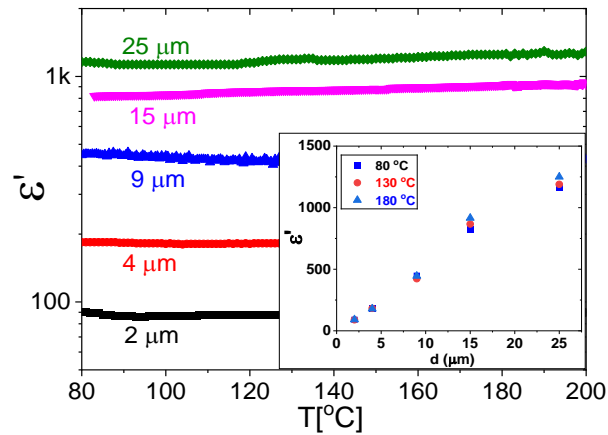

FIG 3. Temperature dependence of dielectric permittivity for planar cells of different thicknesses. Inset: Thickness dependence of apparent dielectric permittivity ε' for planar cells at three different temperatures.

The dielectric permittivities of these cells are temperature independent and linearly dependent on the cell thickness, showing both main features of the extraordinary case ($C_{LC}$ » $C_{al}$) and hence being restricted by Clark's criterion, where the permittivity linearly increases with cell thickness. As a result, it is not possible to accurately determine the real dielectric permittivity in this case.

*2. Dielectric spectroscopy of uncoated cells, Case 1*

To avoid the effect of the capacitance of alignment layers on the dielectric permittivity of the liquid crystal, one needs to increase the capacitance of the alignment layers $C_{al}$ to approach the ordinary case $C_{al}$ » $C_{LC}$, i.e. to increase the ratio $d_{LC}/d_{al}$. The other way to achieve this is simply to use the uncoated (bare) metal/ITO electrodes, as suggested and employed in many studies [11,12,13,14,16,17,25,26,27,28,29]. This removes the restrictions imposed due to the presence of alignment layers and allows us to measure the permittivity up to ~ 300,000 [26].

Therefore, we performed dielectric spectroscopy of WJ-16 in hand-made cells with uncoated ITO electrodes. Figure 4(a, b) shows the 3-D temperature dependencies of the real (ε') and imaginary (ε'') parts of the complex permittivity in the frequency range of 0.1 Hz to 10 MHz in a 4 μm WJ-16 uncoated cell. The complex permittivity data were fitted to the Havriliak-Negami equation as given in Eq. 4. Figure 4(c) illustrates the quality of the dielectric fitting using Eq. 4 applied to three relaxation processes, P1–P3, at a temperature of 130 °C.

$$\varepsilon^* = \varepsilon_\infty + \sum_{j=1}^{n} \frac{\Delta\varepsilon_j}{\left[1+(i\omega\tau_j)^{\alpha_j}\right]^{\beta_j}} - \frac{i\sigma}{\varepsilon_0\omega} \tag{4}$$

Here, $\varepsilon^*$ is the complex dielectric permittivity, $\varepsilon_\infty$ is the high frequency dielectric permittivity that includes both electronic and atomic polarizabilities of the material, ω is the angular frequency of the probe field, $\varepsilon_0$ is the permittivity of free space, $\sigma$ is the DC conductivity, $\tau_j$ is the relaxation time, $\Delta\varepsilon_j$ is the dielectric amplitude or the dielectric strength of the relaxation process, $\alpha_j$ and $\beta_j$ are the symmetric and the asymmetric broadening parameters of the distribution of relaxation times, $j$ in the subscript refers to $j$-th relaxation process.

Let us now compare the results obtained in commercial planar cells with the results of hand-made uncoated cells for the compound WJ-16.

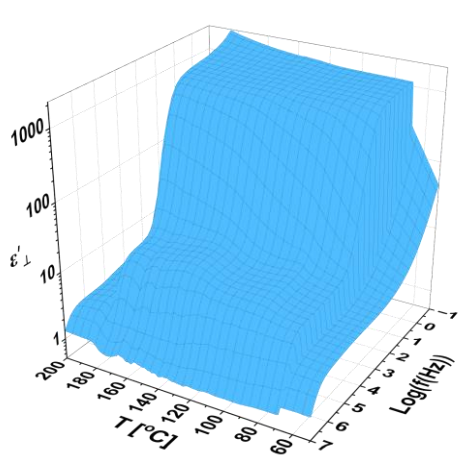
(a)

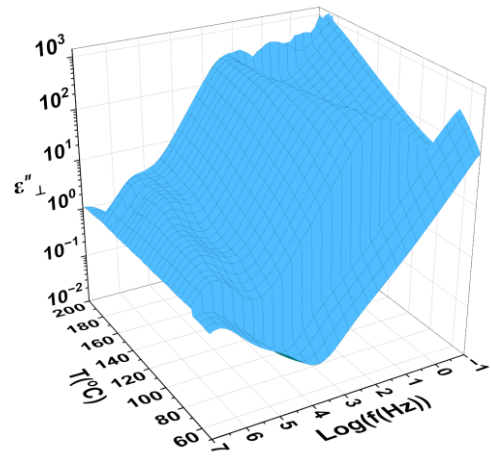
(b)

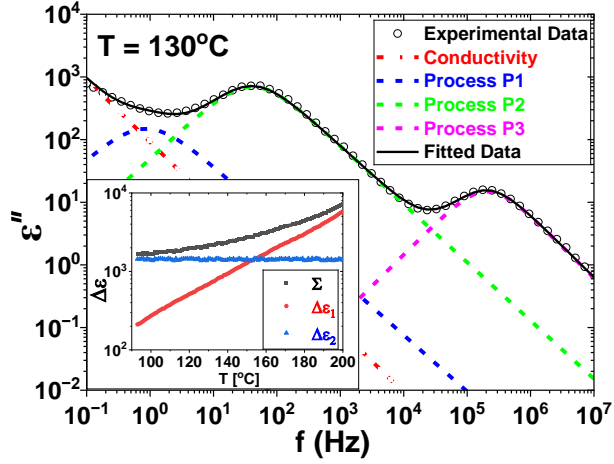

(c)

FIG 4. Temperature dependence of the (a) dielectric permittivity, and (b) loss spectra of 4 μm WJ-16 cell with uncoated electrodes. (c) Example of the dielectric loss spectrum at 130 °C fitted by three relaxation processes: blue (P1), green (P2) and magenta (P3) according to the increase of relaxation frequencies of the processes respectively. Inset: Temperature dependence of dielectric strengths of relaxation processes P1, P2, and their sum.

Upon examining the inset of Fig. 4(c), the observed unusual temperature dependencies of both relaxation processes P1 and P2, immediately capture one's attention. The dielectric strength of the relaxation process P1 ($\Delta\varepsilon_1$) increases exponentially on heating while the dielectric strength of the relaxation process P2 ($\Delta\varepsilon_2$) exhibits colossal permittivity CP (~1400) and is independent of the temperature.

The relaxation processes in WJ-16 are somewhat similar to those observed in DIO where the physical origin of the relaxation processes in DIO were assigned as follows [17,18,19,20]: the lowest-frequency relaxation process P1 arises from the dynamics of ions in the medium having accumulated on the alignment layers [32,33], which is sometimes called as Electrode Polarization (EP) process, and the highest-frequency relaxation process P3 reflects the dynamics of individual molecules around their long molecular axes. The lowest-frequency relaxation process P1 is usually considered as "parasitic" and is responsible for the abnormal rise of the dielectric permittivity on heating to the Isotropic phase.

However, the mid-frequency relaxation process P2 in WJ-16 has a different physical origin than the Goldstone mode in ferroelectric nematics, including DIO. In DIO, the largest P2 is the Goldstone mode, observed only in $N_F$ phase, and was theoretically developed as the so-called polarization-capacitance Goldstone (PCG) mode [24]. However, in WJ-16 it can't be a Goldstone PCG mode because, (1) this material does not have the ferroelectric $N_F$ phase,

(2) this colossal permittivity (CP) is also observed in all phases, including SmA, Nematic and even the Isotropic phase, and (3) its dielectric strength is independent of temperature (see Fig. 4(c)). On the other hand, the total dielectric permittivity is strongly temperature dependent, showing no saturation due to the capacitance of alignment layers. Therefore, we have Case 1. The apparent/ measured value of dielectric permittivity, $\Delta\varepsilon_2 \approx 1400$, is almost seven times higher than the dielectric permittivity of 4 μm planar cell, ~200 (see Fig. 3), and can be accepted as the real value of the perpendicular component of dielectric permittivity, $\Delta\varepsilon_\perp$. Here, it is important to remember that WJ-16 does not have a ferroelectric $N_F$ phase but exhibits CP in a non-ferroelectric superparaelectric (SPE) state [30]. As expected, the dielectric permittivity in SPE is about one order of magnitude lower than in the ferroelectric $N_F$ phase.

### C. Dielectric spectroscopy of the hometropic LC cells, Cases 1 and 3

One of the most important parameters of anisotropic media, such as liquid crystals, is the dielectric anisotropy, $\Delta\varepsilon_a = \Delta\varepsilon_\parallel - \Delta\varepsilon_\perp$, which is responsible for Fréedericksz transition. In the previous section, we showed how to measure the real value of the perpendicular component of dielectric permittivity, $\Delta\varepsilon_\perp$. To measure the parallel component of the dielectric permittivity, $\Delta\varepsilon_\parallel$, one must use a suitable polymer layer for the homeotropic alignment. As done previously for the uncoated cells shown in Fig. 4(c), the dielectric spectra of WJ-16 in 4 μm and 9 μm homeotropic cells were successfully fitted to the Havriliak-Negami equation using three relaxation processes, P1–P3.

Figure 5 shows the temperature dependencies of the dielectric strengths of ionic ($\Delta\varepsilon_1$) and SPE ($\Delta\varepsilon_2$) relaxation processes, P1 and P2, for 4 μm and 9 μm homeotropic cells respectively.

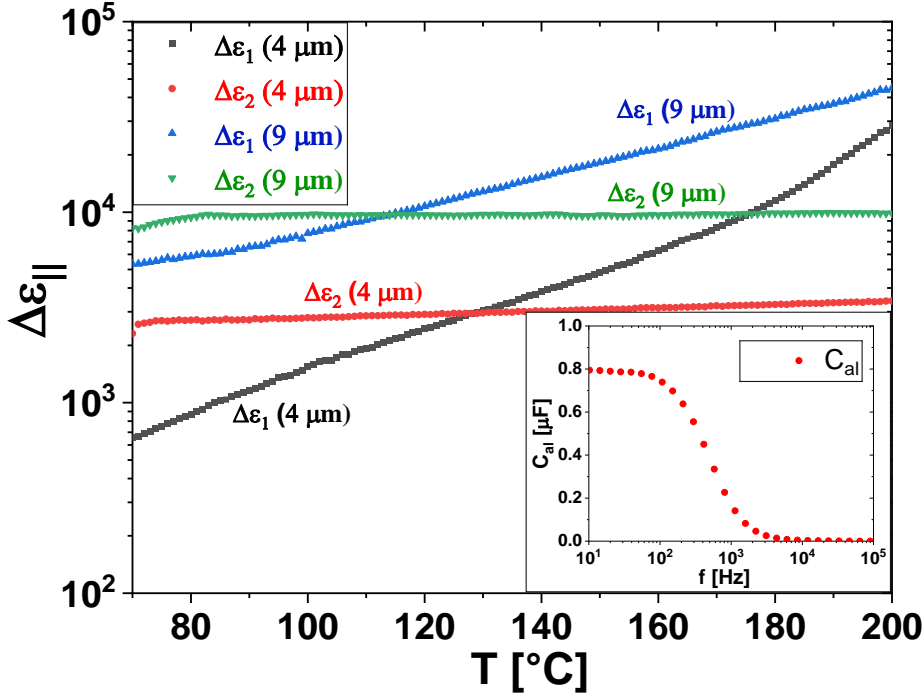

FIG. 5. Temperature dependence of total dielectric strengths of ionic ($\Delta\varepsilon_1$) and SPE ($\Delta\varepsilon_2$) relaxation processes P1 and P2 in 4 μm and 9 μm homeotropic cells. Inset: Frequency dependence of the capacitance of 4 μm commercial cell filled with 20 % NaCl solution in water.

In Fig. 5, one can note that these temperature dependencies are identical to those observed in the uncoated cell (see the inset, Fig. 4(c)), i.e., the dielectric strength of the ionic process ($\Delta\varepsilon_1$) drops exponentially on cooling, while the SPE process ($\Delta\varepsilon_2$) is temperature independent. This suggests that the apparent values correspond to the real dielectric permittivity as in uncoated cells. However, remembering that the electrodes of the homeotropic cells are covered by very thin surfactant (cetyl-trimethyl-ammonium-bromide) monolayers, the measured values might be affected by the capacitance of alignment layers. This scenario can be checked using Eq. (3) for known values of $C_{al}$.

To measure this capacitance, we filled the commercial 4 μm homeotropic cell with a 20% solution of NaCl in water. The frequency dependence of such a cell is shown in the inset of Fig. 5. There is one relaxation process at ~400 Hz due to the mobility of ions. At frequencies ≤ 30 Hz and lower, the cell is shunted by the high conductivity of the water-salt solution, and the measured capacitance ~ 0.8 μF corresponds to the capacitance of alignment layers $C_{al}$. The apparent dielectric permittivity of the SPE relaxation process P2 is ~2800 (see Fig. 5), which corresponds to an apparent capacitance of 1.5 μF. Therefore, here we have an intermediate Case 3, and thereby, on applying Eq. (3), we recover the real value of the liquid crystal layer capacitance as 1.9 μF, which is 24% higher than the apparent value, and in-turn corresponds to the parallel dielectric strength of 3500. Hence, we can calculate the dielectric anisotropy, $\varDelta\varepsilon_a = \varDelta\varepsilon_\parallel - \varDelta\varepsilon_\perp = 3500 - 1400 = 2100$. This is how we may recover the actual value of the parallel component of the dielectric permittivity for a known (measured) value of alignment layer capacitance and thus determine the dielectric anisotropy of the given material.

### *D. Reconstruction of actual value of apparent capacitance, Case 2*

In the previous sections, we have shown some examples of dielectric spectroscopy of ordinary Case 1 in lab-made uncoated cells, extraordinary Case 2 in the commercial planar cells, and intermediate Case 3 in commercial homeotropic cells. Here we now show how the real dielectric permittivity can be recovered even in the extraordinary Case 2.

We performed the dielectric spectroscopy of DIO in the homeotropic cells. The dielectric spectra consist of three relaxation processes, P1-P3, which were assigned as mentioned previously: the lowest-frequency process of P1 corresponds to ion dynamics; mid-frequency process of P2 is a molecular relaxation around short molecular axis, which transfers into the Goldstone mode on transition to the ferroelectric nematic phase, and the highest frequency process P3 reflects the dynamics of individual molecules around their long molecular axes [20].

Figure 6(a) shows the temperature dependence of the dielectric strengths of three relaxation processes, $\varDelta\varepsilon_i$, and the total dielectric strength, $\varDelta\varepsilon = \sum_{i=1}^{3}\varDelta\varepsilon_i$, (which can be considered as the total dielctric permittivity, ε'(ω) at ω → 0, of a 4 μm homeotropic DIO cell [20]. In the mid-temperature range, where the total dielectric permittivity is rather moderate (~< 1000), i.e., well below the limit, all relaxation processes show strong temperature dependence. However, in the upper temperature range, the total dielectric permittivity is rather high (~< 3700). The temperature dependence of the dielectric strengths shows a very weak temperature dependence, indicating that they are rather close to the Clark's criterion. Now, we try to select a limit for the apparent permittivity, which must be reasonably higher than 3700, say 3800, shown as a dashed line in Fig. 6(a).

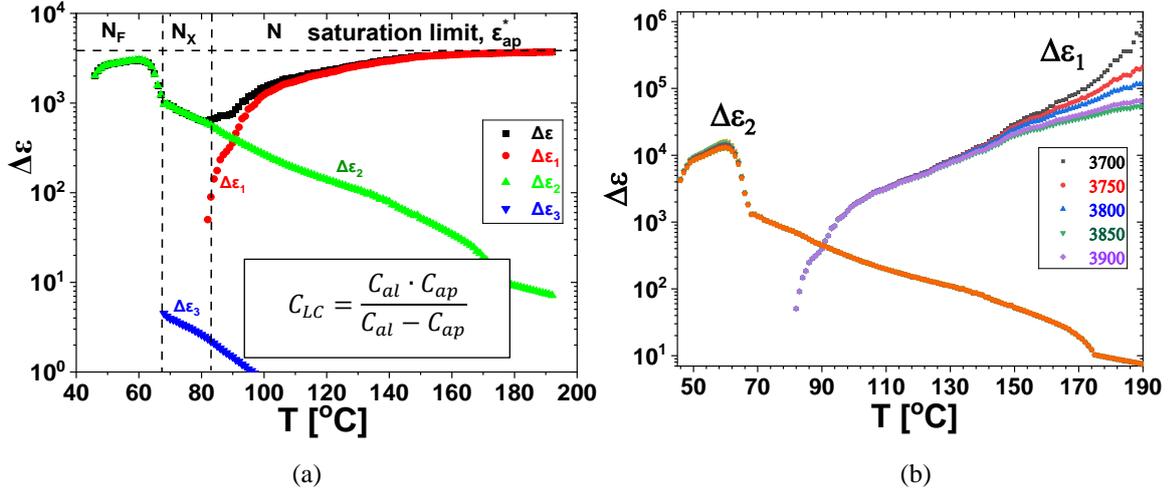

FIG 6. (a) Temperature dependence of the dielectric permittivity in 4 μm homeotropic DIO cell [20]. The data points show the dielectric strengths of different relaxation processes: $\Delta\varepsilon_1$ (●), $\Delta\varepsilon_2$ (▲), $\Delta\varepsilon_3$ (▼) and their sum $\Delta\varepsilon$ (■). The horizontal dashed line at ~3800 is the estimated value of the apparent permittivity limit $\varepsilon_{ap}^*$ assuming full saturation, i.e. $C_{ap} = C_{al}$. (b) Temperature dependence of "recovered" dielectric permittivity of 4 μm DIO homeotropic cell for different apparent value limits from 3700 to 3900 estimated from Fig.1(b) and used as parameter. The data points show the dielectric strengths of two relaxation processes: $\Delta\varepsilon_1$ and $\Delta\varepsilon_2$.

Since the exact value of the limit, $\varepsilon_{ap}^*$, cannot not be precisely defined, we reconstructed the real dielectric permittivity for five different limits in the range $3800 \pm 100$ as a parameter. The temperature dependences of the "recovered" dielectric permittivity of the 4 μm DIO homeotropic cell for different apparent value limits are shown in Fig. 6(b). Noting that the experimental temperature dependence of $\Delta\varepsilon_1$ is exponential and looks like a straight line on the Log Y axis (see Fig. 4 (Inset) and Fig. 5), the best linearity is achieved for the limit value of 3800.

It should be noted that in the $N_F$ phase, where the apparent permittivity ~3200 is rather far from the limit of 3800, the recovered value of the Goldstone mode, $\varepsilon_2$, is less sensitive to selected limit value than the value of the ionic process, $\varepsilon_1$. The recovered value of permittivity ~10,000 is rather realistic and typical for ferroelectric nematic phase. This example shows that the dielectric permittivity can be retrieved even for Case2 with an unknown value of the alignment layer capacitance.

### *E. Further discussions on actual capacitance*

In the previous sections, we discussed the effect of alignment layers on dielectric spectroscopy for three different cases, i.e., ordinary, extraordinary, and intermediate, and demonstrated how the real dielectric permittivity can be recovered from the measured values using Eq. (3).

Now, we shift our attention from the limiting effect of the alignment layers to its interpretation in relation to the apparent (measured) and the real (actual) values of dielectric permittivity, or in other words, are the real values of dielectric permittivity lower or higher than the measured ones? At a first glance, the answer to this question seems obvious. Since this limitation occurs when the capacitance of the liquid crystal exceeds the capacitance of the alignment layers, i.e. $C_{LC} >> C_{al} \approx C_{ap}$, this explicitly implies that $\varepsilon_{LC} >> \varepsilon_{ap}$, i.e., real capacitance of the LC layer and its dielectric permittivity are higher than the apparent value. This also follows directly from Eq. (3).

However, and surprisingly, the original paper on this subject [24] states something quite opposite: "… deriving dielectric constants from electrical impedance measurements of high-polarization ferroelectric liquid crystals, …, can *result in overestimation* of the $\varepsilon'$ values of the LC by many orders of magnitude." [24], i.e. $\varepsilon_{LC} \ll \varepsilon_{ap}$. This is supported by other papers. For example, Ref. [17] states, "Thus, for studied here materials with very high permittivity, the measured equivalent capacity of the circuit might be considerably lower than the actual capacity of LC". Ref. [29] states, "… the relative permittivity of the ferroelectric nematics is indeed huge, and it is even higher than the apparent measured values" and "… measurements confirm a huge relative permittivity of the ferroelectric nematic phase, which can even be orders of magnitude larger than the measured apparent values". However, this contradicts the condition for such a limit, i.e. $C_{LC} >> C_{al} \approx C_{ap}$.

Such curious contradictions can easily be explained by different interpretations of the term "relative permittivity" or, historically "dielectric constant". Ref. [27] states that "In this paper, we have once again *ruled out the CP values of the $N_F$ phase*. In doing so, we have followed the classical definition of permittivity, which *does not include the spontaneous polarization*, as in solid ferroelectrics." Therefore, according to this interpretation, which excludes the ferroelectric Goldstone mode ($\Delta\varepsilon_2$) contributing the largest to the total permittivity, and the rest of the permittivity from other weaker processes ($\Delta\varepsilon_3$) is less than the apparent one. However, such a definition of relative permittivity can hardly be accepted as "classical". According to *classical* definition of "dielectric constant" (or "relative permittivity"), it is defined as the ratio of the total dielectric permittivity to the permittivity of vacuum, $\varepsilon_r(\omega) = \frac{\varepsilon(\omega)}{\varepsilon_0}$, where ε(ω) is the complex frequency-dependent permittivity of the material, which includes <u>*all*</u> relaxation processes regardless of their physical origin and can be measured as $\varepsilon_r(\omega) = \frac{C_{LC}}{C_0}$. Accepting this definition as "classical", the actual dielectric permittivity in ferroelectric nematics is higher than the apparent one. Also, it is not clear from [24,27] whether the authors also exclude Δε$_1$ due to the ionic process P1? If not, as they did not mention about it explicitly, the total dielectric permittivity or capacitance may exceed the capacitance of the alignment layers even without ferroelectric Goldstone mode, exclusively due to ionic mode Δε$_1$, as can be seen in the Fig.6 (a) for T > 150 ºC or due to CP in SPE materials, such as WJ-16, as shown in Fig. 3.

## IV. CONCLUSION

In this paper, we have considered various aspects of dielectric spectroscopy related to the applicability of dielectric measurements for studying ferroelectric nematic liquid crystals. The main attention is paid to the so-called Clark's limit, which is observed in ferroelectric nematics with a huge permittivity, so that the capacitance LC layer can exceed the capacitance of the orienting layers, and the apparent (measured) capacitance will be distorted/limited by the capacitance of the orienting layers, $C_{al}$. We showed that such a limitation is also observed in non-ferroelectric phases such as (i) paraelectric nematic phase due to the high contribution of ionic separation process, and (ii) in new superparaelectric (SPE) LC materials, such as WJ-16 as well. To overcome this limitation, a cell with uncoated ITO layers can be used to measure the perpendicular component of the dielectric permittivity. We also examined different approaches to reconstruct real permittivity values from the apparent values, even in cells with a polymer alignment layer. This is particularly important in retrieving the parallel component of dielectric permittivity in homeotropic cells, which require alignment layers, and hence we determined the dielectric anisotropy, the most important parameter of Liquid Crystals.

Finally, we moderated two opposite interpretations of the effect of alignment layers on apparent dielectric permittivity, showing that it depends on different definitions of dielectric permittivity. According to the "classical" definition, where the permittivity of a material includes all relaxation processes, the actual permittivity is higher than the apparent permittivity. According to another definition of permittivity, which excludes the Goldstone ferroelectric mode, the actual permittivity is lower than the apparent permittivity. Assuming this difference in definitions, both the interpretations are correct.

## ACKNOWLEDGEMENTS

Research work of the Dublin group was funded by the Science Foundation Ireland under the US - Ireland project SFI 21/US/3788.

## 4. References.

[1] C. Filipic, T. Carlson, A. Levstik, B. Zekz, R. Blinc, F. Gouda, S. T. Lagerwall and K. Skarp, Dielectric properties near the smectic-C*–smectic-A phase transition of some ferroelectric liquid-crystalline systems with a very large spontaneous polarization., Phys.Rev.A, **38**, 5833 (1988).

[2] A. M. Biradar, S. S. Bawa, and Subhas Chandra, Dielectric relaxation in a high-tilt-angle chiral-nematic–smectic-C* ferroelectric liquid crystal., Phys.Rev.A, **45**, 7282 (1992).

[3] F. Gouda, W. Kuczynski, S. T. Lagerwall, M. Matuszczyk, T. Matuszczyk and K. Skarp, Determination of the dielectric biaxiality in a chiral smectic-C phase., Phys.Rev.A, **46**, 951 (1992).

[4] Yu. P. Panarin, Yu. P. Kalmykov, S. T. Mac Lughadha, H. Xu and J. K. Vij, Dielectric Response of SSFLC Cells., Phys.Rev.E, **50**, 4763-4772 (1994).

[5] M. Fukui, H. Orihara, A. Suzuki, Y. Ishibashi, Y. Yamada, N. Yamamoto, K. Mori, K. Nakamura, Y. Suzuki, I. Kawamura., Dielectric Dispersion in the Antiferroelectric Liquid Crystal MHPOBC., Jpn.J.Appl.Phys., **29**, L329 (1990).

[6] Yu. P. Panarin, O. E. Kalinovskaya and J. K. Vij, The investigation of the relaxation processes in AFLCs by broad band dielectric and electro-optic spectroscopy., Liq.Cryst., **25**(2), 241-252 (1998).

[7] K. Hiraoka, A. Taguchi, Y Liq.Cryst.,25(2),241-252u. Ouchi, H. Takezoe, A. Fukuda., Jpn.J.Appl.Phys., **29**, L103 (1990).

[8] Yu. P. Panarin, O. E. Kalinovskaya, J. K. Vij and J. W. Goodby., Observation and Investigation of the Ferrielectric Subphase with high qT parameter., *Phys.Rev.E*, **55**, 4345 (1997).

[9] H. Nishikawa, K. Shiroshita, H. Higuchi, Y. Okumura, Y. Haseba, S. Yamamoto, K. Sago and H. Kikuchi, A fluid liquid-crystal material with highly polar order, *Adv. Mater.,* **29**, 1702354 (2017). https://doi.org/10.1002/adma.202101305.

[10] R. J. Mandle, S. J. Cowling and J. W. Goodby, Rational design of rod-like liquid crystals exhibiting two nematic phases, *Chemistry-a European Journal*, **23**, 14554 (2017). https://doi.org/10.1002/chem.201702742.

[11] N. Sebastian, L. Cmok, R. J. Mandle, M. R. de la Fuente, I. Drevenšek Olenik, M. Copic, and A. Mertelj, *Phys.Rev.Lett*. **124**, 037801 (2020).

[12] X. Chen, E. Korblova, D. Dong, X. Wei, R. Shao, L. Radzihovsky, M. Glaser, J. Maclennan, D. Bedrov, D. Walba et al., *Proc. Natl. Acad. Sci. USA*, **117**, 14021 (2020).

[13] J. Li, H. Nishikawa, J. Kougo, J. Zhou, S. Dai, W. Tang, X. Zhao, Y. Hisai, M. Huang, and S. Aya, *Sci. Adv.*, **7**, eabf5047 252627 (2021)

[14] H. Nishikawa and F. Araoka, *Adv. Mater.*, 2101305 (2021)

[15] A. Manabe, M. Bremer, and M. Kraska, *Liq. Cryst.,* **48**, 1079 (2021)
[16] X. Zhao, J. Zhou, H. Nishikawa, J. Li, J. Kougo, Z. Wan, M. Huang, and S. Aya, *Proc. Natl. Acad. Sci.* USA , **118**, e2111101118 (2021)
[17] Brown, E. Cruickshank, J. M. D. Storey, C. T. Imrie, D. Pociecha, M. Majewska, A. Makal, and E. Gorecka, *ChemPhysChem*, **22**, 2506 (2021).
[18] H. Nishikawa, K. Sano, and F. Araoka, *Nat. Comms.*, **13**, 1142 (2022).
[19] N. Yadav, Y. P. Panarin, J. K. Vij, W. Jiang, G. H. Mehl, J. Mol. Liq., **378**, 121570 (2023).
[20] N. Yadav, Yu. P. Panarin, W. Jiang, G.H. Mehl and J. K. Vij, Crystals, **13**, 962 (2023).
[21] J. Hobbs, C. J. Gibb, and R. J. Mandle, Small Sci., 2400189 (2024).
[22] L. Guo, E. Gorecka, D. Pociecha, N. Vaupotič, M. Čepič, R. A. Reddy, K. Gornik, F. Araoka, N. A. Clark, D. M. Walba, K. Ishikawa, H. Takezoe, *Phys. Rev. E,* **84**, 031706 (2011).
[23] S. Nakasugi, S. Kang, Tso-Fu. M. Chang, T. Manaka, H. Ishizaki, M. Sone, J. Watanabe, *J. Phys. Chem. B,* **127***,* 6585−6595 (2023).
[24] N. A. Clark, X. Chen, J.E. Maclennan, M.A. Glaser, *Phys. Rev. Research*, **6**, 013195 (2024).
[25] A. Erkoreka, A. Mertelj, M. Huang, S. Aya, N. Sebastian, and J. Martinez-Perdiguero, *Journal of Chemical Physics*, **159**, 184502 (2023).
[26] A. Erkoreka, J. Martinez-Perdiguero, R. J. Mandle, A. Mertelj, N. Sebastián, *Journal of Molecular Liquids*, **387**, 122566 (2023)
[27] A. Erkoreka and J. Martinez-Perdiguero, *Phys. Rev. E*, **110**, L022701 (2024).
[28] A. Adaka, M. Rajabi , N. Haputhantrige, S. Sprunt, O. D. Lavrentovich , and A. Jákli, *Phys. Rev. Lett.,* **133**, 038101 (2024).
[29] V. Matko, E. Gorecka, D. Pociecha, J. Matraszek and N. Vaupotič, *Arxiv:2401.16084* (2024).
[30] Yu. P. Panarin, W. Jiang, N. Yadav, M. Sahai, Y. Tang, X. Zeng, O. E. Panarina, G. H. Mehl, J. K. Vij, "Colossal Dielectric Permittivity and Superparaelectricity in phenyl pyrimidine based liquid crystals." *J.Mater.Chem. C*, 2024, DOI: 10.1039/D4TC03561E.
[31] N. V. Madhusudana, *Phys Rev. E* 2021, **104**, 014704.
[32] S. Murakami, H. Iga, H. Naito, *J. Appl. Phys.,* **80**, 6396 (1996).
[33] A. Kumar, D. Varshney, J. Prakash, Journal of Molecular Liquids, **303**, 112520 (2020).